\documentstyle[12pt]{article}
\def\be{\begin{equation}}
\def\ee{\end{equation}}
\def\bq{\begin{eqnarray}}
\def\eq{\end{eqnarray}}

\begin{document}

\newpage
\begin{flushright}
CEBAF-TH-94-11
\end{flushright}
\vspace{2cm}
\begin{center}
{\bf Lattice and Continuum Theories.}
\end{center}
\begin{center}{V.M. Belyaev$^*$}
\end{center}
 \begin{center}
 Continuous Electron Beam Accelerator Facility
12000 Jefferson Ave, Newport News, Virginia 23606, USA
\end{center}
\vspace{2cm}
\begin{abstract}
We investigate path integral formalism
for continuum theory. It is shown
that the path integral for the soft modes can be represented in the form of
a lattice theory. Kinetic term of this lattice theory has a standard
form and potential term has additional nonlocal
terms which contributions should tend to zero in the limit
of continuum theory. Contributions of these terms
can be estimated. It is noted that this representation
of path integral may be useful to improve lattice calculations
taking into account hard modes contribution by standard perturbative
expansion.
We discuss translation invariance of correlators and the possibility
to construct a lattice theory which keeps rotary invariance also.

\end{abstract}
\vspace{3cm}
\flushbottom{$^*/\overline{On\;  leave\;  of\;  absence }
\;from\;  ITEP,\;  117259\;  Moscow,\; Russia.$}

\newpage

Path integral formalism \cite{fei} is one of the most useful
tools to study a quantum field
theory.
However there is a serious problem to go out of boundaries of
a perturbative theory.
There are  instanton calculations \cite{inst},
  a   lattice calculation method \cite{lat} and
 variational
approach which can be used in the case of quantum field theory \cite{var}
and sometimes it  is possible  to find nonperturbative exact results
using symmetries of a  quantum field model \cite{susy}.
In ref.\cite{sim} it was used a cluster expansion to take into consideration
nonperturbative effects.

In \cite{l1} it was studied lattice actions which give cut-off independent
physical predictions even on coarse grained lattice.
It was suggested to use {\it perfect} lattice action which is completely
free of lattice artifacts.
 It was shown that
in asymptotical free theories a combination of analytical and numerical
techniques allows to find the perfect action to a sufficient precision.

Here we consider another possibility to improve lattice theory.
In \cite{bel}
it was proposed an alternative method for nonperturbative
 path integral computations.
 All modes are decomposed
 into hard (with $\omega^2 > \omega_0^2$) and soft
(with $\omega^2 < \omega^2_0$) modes where $\omega_0$ is a some parameter.
 It is clear that when a frequency
is enough large then we can consider a potential term as a perturbation
and  use a conventional perturbative theory.  Thus we can find  an effective
Lagrangian \cite{eff} for soft modes using  wellknown  perturbative theory.
Soft modes contribution was estimated by strong coupling expansion.
In \cite{bel} it was shown that this approach is applicable in the case of
quantum mechanics with $V(x)=\lambda x^4$.

Here we show that the path integral for soft modes almost coincides
with a standard lattice definition. To see that let us consider
a path integral
for quantum mechanics:
\bq
<x_f\mid e^{-\hat{H}t_0}\mid x_i>={\cal N}^{-1}\int {\cal D}x(t)
e^{-\int_0^{t_0}{\cal L}(x(t))dt}
\label{1}
\eq
where ${\cal L}(x(t))=\frac{1}{2}(\frac{dx}{dt})^2+V(x)$, $x(0)=x_i$,
$x(t_0)=x_f$, $\hat{H}$ is a hamiltonian of a system, ${\cal N}$ is a
normalization factor.

In the limit
$t_0 \rightarrow \infty$ and with
periodical boundary conditions
$x_i(0)=x_f(t_0)$,  we have
\bq
{\cal Z}=\int dx <x\mid e^{-\hat{H}t_0}\mid x>=\int dx <x\mid n>
e^{-\varepsilon_nt_0}<n\mid x>_{\mid t_0 \rightarrow \infty}
\label{2}
\eq
\bq
=\int dx \mid\Psi_0(x)\mid^2e^{-\varepsilon_0t_0}=e^{-\varepsilon_0t_0}
\nonumber
\eq
where $\varepsilon_n$ is the energy of the $n-$th state, and $\varepsilon_0$ is
the lowest energy of the system.
The factor ${\cal N}$ is chosen in the following form:
$\int {\cal D}x(t)e^{-\int_0^{t_0}\frac{1}{2}(\frac{dx}{dt})^2dt}$.

In a perturbative theory the following basis for trajectories  is  used
\bq
x(t)=\sum_{n=-\infty}^{+\infty}C_ne_n(t)
\label{3}
\eq
where $e_n(t)=\frac{1}{\sqrt{t_0}}e^{i\omega_nt}$,
$\omega_n=\frac{2\pi}{t_0}n$,
$C_n=C_{-n}^*$.

This basis $\{ e_n\}$  has the  normalization:
$
<e_n\mid e_m> = <e_n^*e_m>=\int_0^{t_0}e^*_n(t)e_m(t)dt=\delta_{mn}
$
and  in the basis (\ref{3}) path integral has the following form
\bq
{\cal Z}={\cal N}^{-1}\int \prod_{n=-\infty}^{+\infty}\frac{dC_n}{\sqrt{2\pi}}
e^{-<{\cal L}(\sum_nC_ne_n)>}
\label{4}
\eq
Here we use the denotation: $<f(t)>=\int_0^{t_0}f(t)dt$.

Hard modes can be taken into consideration by conventional perturbative
procedure and  after  integration over them we obtain a low energy
effective Lagrangian for the soft modes.

In \cite{bel} another basis for
trajectories  was suggested:
\bq x(t)=\sum_{\mid n \mid < N}B_nE_n(t)+\sum_{\mid n\mid >N}C_ne_n(t)
\label{5} \eq \bq \omega_0=\frac{2\pi}{t_0}N;\;\;
\nonumber
\eq

Where
$
E_n(t)=\frac{1}{\sqrt{\Delta t}}\Theta (t-t_0/2-n\Delta t)\Theta
(t_0/2+(n+1)\Delta t -t)
$
and $\Delta t=\pi/\omega_0$.

The following denotations are used below:
greek letters: $\mu$, $\nu$,..$=0,\pm 1,..,\pm N$;
small letters: $m$, $n$,..$=\pm(N+1),\pm(N+2),..$;
large letters: $M$, $L$,..$=0,\pm 1,..\;\infty$

In \cite{bel} it was shown that
\bq
{\cal Z}={\cal N}^{-1}\int \prod_n\frac{dC_n}{\sqrt{2\pi}}\prod_\mu
\frac{dB_\mu}{\sqrt{2\pi}}
e^{-<{\cal L}(\sum (C_ne_n+B_\mu (E_\mu -<E_\mu e^*_n>e_n) ))>}\mid J \mid
\label{8}
\eq
where
$
J=det ( <e_\mu E_\nu >)=e^{-\frac{\omega_0 t_0}{\pi}j}
$
where $j=\ln (\pi )-1\simeq 0.14$.
In quantum mechanics lagrangian has a form
\be
{\cal L}=\frac{1}{2}(\frac{dx}{dt})^2+V(x)
\label{b1}
\ee
Then in the terms of our basis $\{ E_\mu\}+\{ e_n\}$ the path integral is
\be
{\cal Z}=\frac{1}{\cal{N}}\int \prod_n\frac{dC_n}{\sqrt{2\pi}}\prod_\mu
\frac{dB_\mu}{\sqrt{2\pi}}|J|
\label{b3}
\ee
\bq
\times
\exp \{ -[\frac{1}{2}|C_n|^2\omega_n^2+
\frac{1}{2}B_\mu <E_\mu e_\rho>\omega_\rho^2<e_\rho^* E_\nu>B_\nu
\nonumber
\eq
\bq
+
<V(B_\mu (E_\mu -<E_\mu e_n>e_n)>
+(terms\; with\;C_n)]\}
\nonumber
\eq
It is easy to show that the kinetic term for soft modes
coincides with lattice definition:
$\frac{1}{2}\frac{(x_{\mu+1}-x_\mu )^2}{\Delta}$ where $\Delta=
\frac{\pi}{\omega_0}$ is the lattice size.

Potential for soft modes $B_\mu$ is
\bq
<V(B_\mu (E_\mu -<E_\mu e^*_n>e_n)>
\label{wp}
\eq
\bq
=
<V(B_\mu E_\mu )>-<(\frac{d}{dx}V(x)_{\mid x=B_\mu E_\mu}e_n>
<e^*_n E_\nu >B_\nu +...
\nonumber
\eq
Due to the definition $E_\nu (t)$ the first term in the expansion
of eq.(\ref{wp}) coincides with lattice definition. The rest terms
of the expansion in (\ref{wp}) are not presented in the lattice
formulation of the path integral. So, suggesting that lattice
theory and continuum formulation of the path integral describe
the same system in the limit $\delta\rightarrow 0$, we have to
conclude that in this limit ($\omega_0\rightarrow\infty$) the
contribution of the highest terms of expansion (\ref{wp}) should tend to
zero and at a finite $\omega_0$ we can estimate them.
In \cite{bel} it was shown that correction
of this nonlocal terms is about few percent at $\omega_0\sim\lambda^{1/3}$
in quantum mechanics with $V(x)=\lambda x^4$.

Let us consider the soft modes contribution to correlator $<x(t_1),x(t_2)>$:
\bq
<x(t_1),x(t_2)>=
\label{cor}
\eq
\bq
=<B_\mu (E_\mu (t_1)-<E_\mu e_m^*>e_m(t_1)),
B_\nu (E_\nu (t_2)-<E_\nu e_n^*>e_n(t_2))>
\nonumber
\eq
\bq
=e_\mu (t_1)<e_\mu^* E_\rho ><B_\rho B_\lambda ><E_\lambda e_\nu^*>
e_\nu (t_2)=e_\mu (t_1)W_{\mu\nu}e_\nu (t_2)
\nonumber
\eq
Using that $<B_\rho B_\lambda >$ is a periodical function over
$\rho$ and $\lambda$ it is possible to show that
$W_{\mu\nu}\sim\delta_{\mu,-\nu}$. Thus we see that the correlator
depends on $(t_1-t_2)$ only. We see that translational invariance is
not broken if we use any approximation of the path integral keeping
periodical boundary conditions. It is true in the case of any correlators
and it is possible to use this definition of correlators (\ref{cor})
in lattice calculations.

The formalism suggested here can be expanded to the case of
$d-$dimensional scalar field theory directly using cubic lattice
and the following definition for soft modes: $\mid p_i\mid<\omega_0$,
$p_i$ is one of components of particle momentum.
The main features of the approach are: 1. Soft modes contribution can be
calculated by lattice computations with additional nonlocal terms which
should tend to zero at $\omega_0\rightarrow\infty$; 2. It is possible to
improve lattice results using effective lagrangian for soft modes; 3. All
renormalizations are taking into account in effective lagrangian by a
standard way; 4. Translational invariance is not broken in any approximation
keeping periodical boundary conditions.  Notice that rotary symmetry of
quantum field theory is broken in this approach. It should be restored due
to the hard modes contribution.
It is possible to construct lattice theory which keeps rotary invariance.
To obtain this kind of lattice theory it is enough to use cubic lattice for
soft modes and rotational invariant basis for hard ones.
It is  possible to see that in this case kinetic term does not coincide
with standard lattice term and it would be interesting to compare this
kinetic term with the perfect action of \cite{l1}.
 The case of gauge theory should be
investigated separately to try to find the way keeping gauge invariance.

Author would like to thank A.A.Abrikosov(jr.), A.Lossev and Yu.A.Simonov
for fruitful discussions.


\begin{thebibliography} {99}
\bibitem{fei}R.P. Feynman, A.R. Hibbs, {\it Quantum mechanics and
Path Integrals} McGraw-Hill, New York (1965).
\bibitem{inst}A.A. Belavin, A.M. Polyakov, A.S. Shvarts, Yu.S. Tyupkin,
Phys.Lett. 59B (1975) 85; A.A. Belavin, A.M. Polyakov, Nucl.Phys.
B123 (1977) 429.
\bibitem{lat}R. Balian, J.M. Drouffe, C. Itzykson, Phys.Rev. D11 (1975)
2104; J. Banks, L. Susskind, J. Kogut, Phys.Rev. D13 (1976) 1043.
\bibitem{var}D.I. Diakonov, V.Yu. Petrov, Nucl.Phys. B245 (1984) 259.
\bibitem{susy}A.I. Vainshtein, V.I. Zakharov,  M.A. Shifman,
JETP Lett. 42 (1985) 224.
\bibitem{sim}H.G. Dosch, Yu.A.Simonov, Phys.Lett. 205B (1989) 339.
\bibitem{l1}P. Hasenfratz and F. Niedermayer, Nucl.Phys. B414 (1993) 785.
\bibitem{bel}V.M.Belyaev, Int.J.Mod.Phys. A8 (1993) 4019.
\bibitem{eff}S. Coleman, S. Weinberg, Phys.Rev. D7 (1973) 1888.

\end{thebibliography}
 \end{document}